\documentclass[%
 reprint,superscriptaddress,
 amsmath,amssymb,pra]{revtex4-1}

\usepackage{physics}
\usepackage{xcolor}
\usepackage{xtab,afterpage,longtable}
\usepackage{lipsum}
\usepackage{graphicx}
\usepackage{dcolumn}
\usepackage{booktabs} 
\usepackage{multirow}
\usepackage{color}
\newcommand*{\rom}[1]{\expandafter\@slowromancap\romannumeral #1@}

\begin{document}

\title{Communication-efficient Quantum Algorithm for Distributed Machine Learning}

\author{Hao Tang}
\thanks{These authors contributed equally.}
\affiliation{Department of Materials Science and Engineering, Massachusetts Institute of Technology, MA 02139, USA}

\author{Boning Li}
\thanks{These authors contributed equally.}
\affiliation{
   Research Laboratory of Electronics, Massachusetts Institute of Technology, Cambridge, MA 02139, USA}
\affiliation{Department of Physics, Massachusetts Institute of Technology, MA 02139, USA}

\author{Guoqing Wang}%
\affiliation{
   Research Laboratory of Electronics, Massachusetts Institute of Technology, Cambridge, MA 02139, USA}
\affiliation{
   Department of Nuclear Science and Engineering, Massachusetts Institute of Technology, Cambridge, MA 02139, USA}
\author{Haowei Xu}%
\affiliation{
   Department of Nuclear Science and Engineering, Massachusetts Institute of Technology, Cambridge, MA 02139, USA}
\author{Changhao Li}%
\affiliation{
   Research Laboratory of Electronics, Massachusetts Institute of Technology, Cambridge, MA 02139, USA}
\author{Ariel Barr}%
\affiliation{Department of Materials Science and Engineering, Massachusetts Institute of Technology, MA 02139, USA}
\author{Paola Cappellaro}%
 \email{pcappell@mit.edu}
 \affiliation{
   Research Laboratory of Electronics, Massachusetts Institute of Technology, Cambridge, MA 02139, USA}
 \affiliation{Department of Physics, Massachusetts Institute of Technology, MA 02139, USA}
\affiliation{
   Department of Nuclear Science and Engineering, Massachusetts Institute of Technology, Cambridge, MA 02139, USA}
   
\author{Ju Li}%
 \email{liju@mit.edu}
 \affiliation{Department of Materials Science and Engineering, Massachusetts Institute of Technology, MA 02139, USA}
\affiliation{
   Department of Nuclear Science and Engineering, Massachusetts Institute of Technology, Cambridge, MA 02139, USA}
   
\date{\today}
             
\begin{abstract}
The growing demands of remote detection and increasing amount of training data make distributed machine learning under communication constraints a critical issue. 
This work provides a communication-efficient quantum algorithm that tackles two traditional machine learning problems, the least-square fitting and softmax regression problem, in the scenario where the data set is distributed across two parties. 
Our quantum algorithm  finds the  model parameters with a communication complexity of $O(\frac{\log_2(N)}{\epsilon})$, where $N$ is the number of data points and $\epsilon$ is the bound on parameter errors. Compared to classical algorithms and other quantum algorithms that achieve the same output task, our algorithm provides a communication advantage in the scaling with the data volume. The building block of our algorithm, the quantum-accelerated estimation of distributed inner product and Hamming distance, could be further applied to various tasks in distributed machine learning to accelerate communication. 

\end{abstract}

\maketitle

The amount of training data is critical for machine learning models to achieve high accuracy, generalization capabilities and prediction power. 
At the same time, the total amount of stored data worldwide is growing with unprecedented speed, so it becomes a challenge for  machine learning algorithms to exploit such large-scale data within feasible time and memory~\cite{gheisari2017survey,bottou2007tradeoffs}. 
Distributed machine learning emerges as a promising solution, where the training data and learning process are allocated among multiple machines~\cite{verbraeken2020survey,peteiro2013survey}. 
Distributed algorithms naturally scale up computational power and also provide a way to deal with intrinsically distributed data when collected~\cite{erickson2009database}. However, these algorithms require extensive communication between different machines, which  usually becomes a rate-limiting step~\cite{li2017fundamental}. 
Therefore, efficient communication schemes for machine learning tasks are attracting broad interest. 
The necessary communication between two machines in a computation task is quantified by its communication complexity, 
either within classical~\cite{abelson1980lower,yao1979some,kushilevitz1997communication,rao2020communication} or quantum channels~\cite{brassard2002quantum,martinez2018high,brassard2003quantum,buhrman1998quantum}.
Even though quantum algorithm have been shown to  reduce the communication complexity compared to classical communication in various scenarios~\cite{buhrman2010nonlocality}, these do not include the field of machine learning, where instead quantum algorithms have been studied so far only as accelerators for the computational complexity~\cite{biamonte2017quantum}. Square-root  or exponential speedups have been  demonstrated in many problems, such as least-square fitting~\cite{wiebe2012quantum}, statistical inference~\cite{low2014quantum}, feature engineering~\cite{lloyd2014quantum}, and classification problems~\cite{rebentrost2014quantum}. In comparison, whether quantum algorithms can accelerate communication in distributed learning tasks remains an open question. 

Here, we propose a quantum communication algorithm for two typical data fitting subroutines in machine learning: least-square fitting and softmax regression, which are common output layers of predictors and classifiers, respectively~\cite{lecun2015deep}. 

A typical training dataset contains $N$ independent identically distributed (\textit{iid}) data points. Each data point has an $M$-dimensional input $\vec{x}$ and a scalar output $y$. 
In the basic communication scenario~\cite{peteiro2013survey}, the training dataset, comprising  the input attributes and labels, is distributed across two parties, Alice and Bob. 
Both least-square fitting and softmax regression aim at
fitting a model $y \approx f(\vec{x},\boldsymbol{\lambda})$ to the data,  by estimating the parameters $\hat{\boldsymbol{\lambda}}$ that minimize a given loss function.
The goal of a communication algorithm is to minimize  the number of bits~\cite{abelson1980lower,yao1979some} or qubits~\cite{brassard2003quantum,buhrman1998quantum} exchanged between Alice and Bob during model fitting, while keeping the accuracy of $\hat{\boldsymbol{\lambda}}$ within a standard error $\epsilon$.

Least-square fitting has been extensively studied in both classical distributed algorithms and single-party (no communication) quantum algorithms. Using a classical algorithm based on correlation estimation, it has been proved that the classical communication complexity cannot be below $O(1/\epsilon^2)$~\cite{hadar2019communication,freedman2009statistical}.
However, to reach such lower bound requires an exponentially large number of data points. In the case of finite datasets, since the accuracy of the fitting parameters should be at least  as small as its error $\epsilon$, a classical deterministic method  requires $O(N\log_2({1/\epsilon}))$ bits to be  exchanged between two parties within a precision $\epsilon$~\cite{burden2015numerical}. When  high accuracy is not required,  only $1/{\epsilon^2}$ data points with random indexes  need to be transferred, which yields a $O((\log_2(1/\epsilon)+\log_2(N))\frac{1}{\epsilon^2})$ communication complexity~\cite{hadar2019communication}. Then, to achieve a statistical variance $\epsilon_{s}^2 = var(|\lambda|)\propto 1/{N}$, these two classical algorithms have the same communication complexity $O(N\log_2(N))$ or $O(\frac{\log_{2}(1/\epsilon_{s})}{\epsilon_{s}^2})$.
In comparison, quantum computation methods for linear fitting based on the Harrow-Hassidim-Lloyd (HHL) algorithm~\cite{harrow2009quantum} yield a quantum state $|\lambda\rangle = \sum_{j=1}^M \lambda_j|j\rangle$ encoding the fitting parameters in the superposition amplitudes with communication complexity of $O(\log_2(N))$~\cite{wiebe2012quantum,zhang2019realizing,schuld2016prediction}.
 However, these results have practical limitations. As the quantum state only encodes normalized parameters ($|\boldsymbol{\lambda}|^2=1$),  additional computation resource are needed to obtain the real value. In addition, to extract the (normalized) parameters $\lambda_{j=1..M}$, the HHL-based algorithm requires $O(M^2\frac{1}{\epsilon^2})$  repeated measurements, which is inefficient when a small error $\epsilon$ is required. 
 Within the communication scenario, the HHL-based fitting algorithm requires communicating $O(\frac{\log_2(N) }{\epsilon^2})$ qubits  to determine $\lambda_j$~\cite{wiebe2012quantum,wang2017quantum}, with no clear advantage over classical algorithms.

We designed a \textit{quantum counting}-based~\cite{counting,nielsen2001quantum} communication algorithm that achieves a reduced communication complexity of $O(\frac{\log_2(N)}{\epsilon})$ for both least-square fitting and softmax regression (Table~\ref{table:complexity}). 
At its core, the direct action of our algorithm is to estimate the correlation or the Hamming distance of  two bit-strings  distributed across two parties.  
Embedding this algorithm into a hybrid computing scheme enables the data fitting tasks beyond the theoretical limit of classical algorithms, and we expect it could benefit other scenarios not analyzed here.

\begin{table*}
\renewcommand\arraystretch{1.4}
\caption{Comparison of the communication complexity of classical distributed algorithm, quantum counting-based algorithm developed in this work, and other quantum algorithms. Listed problems include estimating correlation and Hamming distance of two separate bit strings, distributed linear fitting, and distributed softmax regression. In the first column, (c) and (q) means the problem requires output as classical data or quantum states, respectively. In the table, $\epsilon$, $N$, $M$ are the standard error of solution, number of data points, and number of attributes in Alice's data; $\kappa$ and $s$ are the condition number and sparseness of the matrix $\boldsymbol{X}$ in linear regression problems; and $q$ is the number of classes in softmax regression problems. (See derivation of the terms in the table in section~\uppercase\expandafter{\romannumeral3}.).}
\begin{tabular}{ p{3.5cm}p{4.5cm}p{3cm}p{5.8cm}  }
 \hline
 \hline
Problem (output)  & Classical algorithm & Quantum counting & Other quantum algorithm \\
 \hline
 Correlation (c)    & $O(\frac{1}{\epsilon^2})^a$(lower-bound) & $O(\frac{\log_2{(N)}}{\epsilon})$ & $O(\frac{\log_2{(N)}}{\epsilon^2})$  (swap-test)$^b$   \\
\hline
 Hamming distance (c)    & $O(N)^d$  &  $O(\frac{\log_2(N)}{\epsilon})$ & $--$\\
\hline
 Linear-fitting (c)     & $O(N\log_2{(\frac{\kappa^2}{\epsilon})})$(deterministic)$^g$ $O(\frac{\log_2(N)+\log_2{(\frac{\kappa^2}{\epsilon})}}{(\epsilon/\kappa^2)^2})$(stochastic)$^a$ &  $O(M\kappa\frac{\log_2(N)}{\epsilon})$ & $O(M^2\kappa^5\frac{\log_2(N)}{\epsilon^2})$ (HHL)$^{e,f}$\\
\hline
 Linear-fitting (q)    & $--$  &  $O(M\kappa\frac{\log_2(N)}{\epsilon})$ & $O(\kappa^5\log_2(N))$ (HHL)$^e$\\
\hline
 Softmax regression (c)    & $O(N\log_2{q})$ & $O(Mq\kappa \frac{\log_2(N)}{\epsilon})$  & $--$\\
 \hline
 \hline
\end{tabular}
\footnotetext[1]{Ref.~\cite{hadar2019communication}, $^b$ Ref.~\cite{fanizza2020beyond}, $^c$ Ref.~\cite{anshu2022distributed},  $^d$ Ref.~\cite{chakrabarti2012optimal}, $^e$ Ref.~\cite{wiebe2012quantum}, $^f$ Ref.~\cite{wang2017quantum}, $^g$ Ref.~\cite{burden2015numerical}}
\label{table:complexity}
\end{table*}

\begin{figure}[t]
\centering
\includegraphics[width=\linewidth]{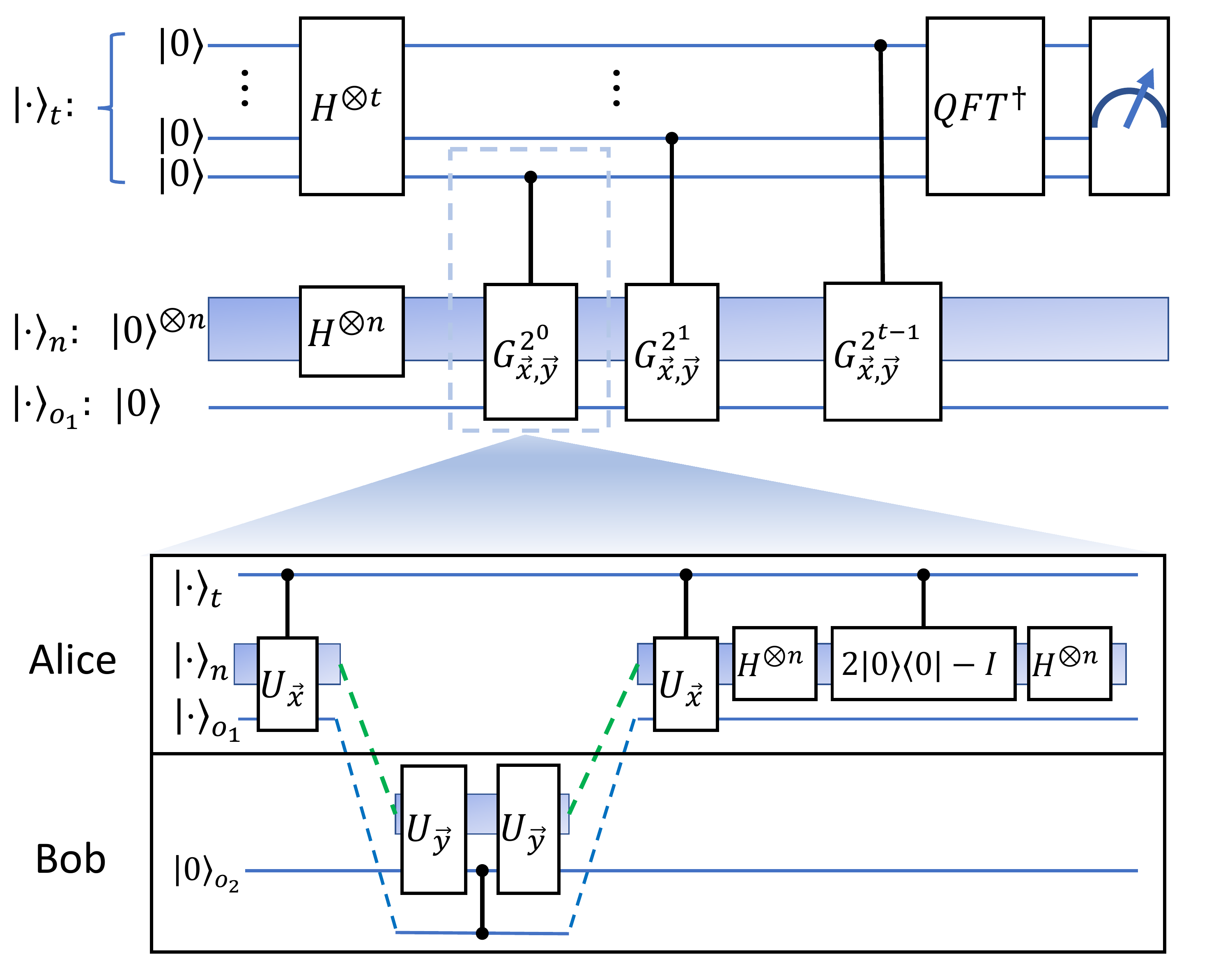}
\caption{Quantum circuits for the distributed quantum counting scheme. 
$H$, $G$, and $QFT^{\dagger}$ represent the Hadamard gate, the Grover operator, and the inverse QFT, respectively. The $t$-qubit register is measured after the inverse QFT. 
The inset shows the biparty scheme of the Grover operation, where $U_{x_l}$ and $V_{y_k}$ are defined in Eq.~(7,8).}
\label{fig:circuit}
\end{figure}

\paragraph*{Estimating correlation.}
We first present the core subroutine of our algorithm, the quantum counting-based communication scheme for the inner product. The problem is stated as follow: Alice and Bob have $N$-dimensional  vectors $\vec{x}^{\rm b}, \vec{y}^{\rm b}\in \{0,1\}^N$, respectively, that can only take binary values (denoted by superscript $^{\rm b}$). This is not restrictive, as  real numbers can always be expanded as binary floating point numbers 
(see section "least-square fitting"). The task is to estimate the correlation $\hat{\rho}\equiv\frac{\overline{x^{\rm b}y^{\rm b}}-\overline{x^{\rm b}}\cdot\overline{y^{\rm b}}}{\sqrt{\overline{x^{\rm b}}(1-\overline{x^{\rm b}})\overline{y^{\rm b}}(1-\overline{y^{\rm b}})}}$, in which the communication-intensive step is to evaluate $ \overline{x^{\rm b}y^{\rm b}}=\frac{1}{N}\sum_{i=1}^{N}x^{\rm b}_iy^{\rm b}_i$ within a standard deviation error $\epsilon$~\cite{hadar2019communication}. 
 
We assume that Alice and Bob have access to quantum computers with oracles.
The oracle of Alice's computer performs a unitary transformation $\hat{U}_{\vec{x}^{\rm b}}^{1,2}: |i\rangle_1 |0\rangle_2 \mapsto |i\rangle_1 |x^{\rm b}_i\rangle_2$ that encodes the data $x^{\rm b}_i$, where $|i\rangle$ is an $n\equiv\lceil\log_2(N)\rceil$-qubit state $|i_1i_2\cdots i_n\rangle$, representing the index of the queried component, and $|x^{\rm b}_i\rangle$ is a single-qubit state. 
Bob has an oracle $\hat{U}_{\vec{y}^{\rm b}}$ of the same type that encodes the data $y^{\rm b}_i$. This type of oracle is a common building block in quantum algorithms~\cite{wiebe2011simulating,wiebe2012quantum,harrow2009quantum}, 
which can be realized through quantum random access memory~\cite{giovannetti2008quantum} or other data loading procedures~\cite{zhang2021low,cortese2018loading}.

Estimating the correlation $\overline{x^{\rm b}y^{\rm b}}$ is based on the quantum counting algorithm, in which the phase oracle is realized cooperatively by Alice and Bob through communication, as shown in Fig.~\ref{fig:circuit}. We sketch the framework here and provide the algorithm details in the supplementary materials (SM) section~\uppercase\expandafter{\romannumeral1}. The algorithm works on an $n$-qubit vector index space ($|\cdot\rangle_n$), a $t$-qubit register space ($|\cdot\rangle_t$), and a 2-qubit oracle workspace ($|\cdot\rangle_o$). Initially, all qubits are set to zero: $\ket{\psi_0} \equiv |0\rangle_t|0\rangle_n|00\rangle_o$. 
Hadamard gates are applied to create superposition in both $t$ and $n$ space $\ket{\psi_1} = 2^{-(t+n)/2}\sum_{i,\tau}|\tau\rangle_t|i\rangle_n|00\rangle_o$. A phase oracle on the state $|\cdot\rangle_n$ can be realized through the following unitary operation:
\begin{equation}
\hat{O}_{\vec{x}^{\rm b},\vec{y}^{\rm b}}\equiv \hat{U}_{\vec{x}^{\rm b}}^{n,o_1}\hat{U}_{\vec{y}^{\rm b}}^{n,o_2}CZ^{o_1,o_2}\hat{U}_{\vec{y}^{\rm b}}^{n,o_2}\hat{U}_{\vec{x}^{\rm b}}^{n,o_1},
\label{eq:oracle}
\end{equation}
which yields $\hat{O}_{\vec{x}^{\rm b},\vec{y}^{\rm b}}|i\rangle_n|00\rangle_o = (-1)^{x^{\rm b}_iy^{\rm b}_i}|i\rangle_n|00\rangle_o $. Here $o_1,o_2$ are the two qubits in the oracle space and $CZ^{o_1,o_2}$ is a control-Z gate acting on them. 
Each oracle call requires about $2n$-qubit communication, as Alice needs to send the $(n+1)$-qubits to Bob after applying $\hat{U}_{\vec{x}^{\rm b}}^{n,o_1}$ and Bob needs to send the $(n+1)$-qubits back after applying $\hat{U}_{\vec{y}^{\rm b}}^{n,o_2}CZ^{o1,o2}\hat{U}_{\vec{y}^{\rm b}}^{n,o_2}$; finally, Alice applies $\hat{U}_{\vec{x}^{\rm b}}^{n,o_1}$ to finish the whole oracle $\hat{O}_{\vec{x}^{\rm b},\vec{y}^{\rm b}}$. The Grover operation needed for counting is then constructed as $\hat{G}_{\vec{x}^{\rm b},\vec{y}^{\rm b}} \equiv\hat{H}^{\otimes n} (2|0\rangle_n \langle 0|_n-\hat{I})\hat{H}^{\otimes n}\hat{O}_{\vec{x}^{\rm b},\vec{y}^{\rm b}}$. The quantum counting scheme applies the Grover operation iteratively on the initial state:
\begin{equation}
|\psi_2\rangle = \frac{1}{2^{(t+n)/2}}\sum_{\tau} |\tau \rangle_t \otimes (\hat{G}_{\vec{x}^{\rm b},\vec{y}^{\rm b}})^{\tau}\sum_i |i\rangle_n|00\rangle_o.
\label{eq:grover}
\end{equation}
Expanding the Grover operator in its eigenbasis gives  
$(\hat{G}_{\vec{x}^{\rm b},\vec{y}^{\rm b}})^{\tau}\sum_{i}\ket{i}_n = (e^{i\tau\theta}|\phi_+\rangle\langle \phi_+|+e^{-i\tau\theta}|\phi_-\rangle\langle \phi_-|)\sum_{i}\ket{i}_n$,
where $|\phi_{\pm}\rangle$ are the two eigenstates of  $\hat{G}_{\vec{x}^{\rm b},\vec{y}^{\rm b}}$, and $\theta = 2 \arcsin(\sqrt{\overline{x^{\rm b}y^{\rm b}}})$.
Applying the inverse quantum Fourier transform ${\rm QFT}^{\dagger}$ to  $|\cdot\rangle_t$ yields the final state:
\begin{equation}
\ket{\psi_3}\!=\!\frac{1}{\sqrt{2^{t+n}}}\!\!\sum_{\eta=\pm, i}\!\!\langle \phi_{\eta}|i\rangle |\phi_{\eta}\rangle_n|00\rangle_o {\rm QFT}^{\dagger}(\sum_{\tau} |\tau \rangle_t e^{i\eta\tau\theta}).
\label{eq:counting}
\end{equation}
Measuring the t-register will project into a state $|j\rangle_t$ resulting in the phase $2\pi j\cdot2^{-t}$ which encodes either $\hat{\theta}$ or $2\pi-\hat{\theta}$ with equivalent standard deviation: $\Delta \hat{\theta} = 2^{-t+1}$.

Both cases  give the same estimated correlation $\widehat{\overline{x^{\rm b}y^{\rm b}}} = \sin^2(\frac{\hat{\theta}}{2})$, with standard deviation $\epsilon = \sqrt{\overline{x^{\rm b}y^{\rm b}} (1-\overline{x^{\rm b}y^{\rm b}})}2^{-t+1}$ (see SM section~\uppercase\expandafter{\romannumeral2} for details). 
The overall communication complexity $\mathcal C$ is the Grover operation's $2(n+1)$ qubits communication repeated for $2^t-1$ iterations: 
\begin{equation}
	\mathcal C = 2(n+1)(2^t-1) = O\left(\frac{\log_2(N)}{\epsilon}\right),
	\label{eq:communication}
\end{equation}
where we choose $t$ to satisfy the desired error bound.
The computational complexity is the total number of oracle calls by Alice and Bob, which is $\mathcal C_{\rm comp} = 4(2^t-1)=O(\frac{1}{\epsilon})$.

We note that our algorithm solves the problem of estimating $\overline{x^{\rm b}y^{\rm b}}$,  which is  equivalent to computing the inner product. Inner product of quantum states is usually accomplished by the swap test algorithm~\cite{fanizza2020beyond,anshu2022distributed}. However, the swap test method costs $O(\frac{\log_2(N)}{\epsilon^2})$ bits of communication, due to the requirement of repeated measurements. Recently, A. Anshu, et al~\cite{anshu2022distributed} proposed an algorithm to estimate the inner product of two quantum states using local quantum operations and classical communication (LOCC). With respect to communication complexity, neither of the algorithms achieves an advantage over the classical algorithms.

\paragraph*{Estimating the Hamming distance.}
The  algorithm can also estimate the Hamming distance $d$ between $\vec{x}^{\rm b}$ and $\vec{y}^{\rm b}$ (that is, the number of positions $i$ where $x^{\rm b}_i \neq y^{\rm b}_i$). The key is to replace the oracle in Eq.~\ref{eq:oracle} by
\begin{equation}
\hat{O}'_{\vec{x}^{\rm b},\vec{y}^{\rm b}}\equiv \hat{U}_{\vec{x}^{\rm b}}^{n,o_1}\hat{U}_{\vec{y}^{\rm b}}^{n,o_2}C_{NOT}^{o_1,o_2}Z^{o_2}C_{NOT}^{o_1,o_2}\hat{U}_{\vec{y}^{\rm b}}^{n,o_2}\hat{U}_{\vec{x}^{\rm b}}^{n,o_1},
\label{eq:hamming}    
\end{equation}
where $C_{NOT}^{o_1,o_2}$ represents a CNOT gate with $o_1$ as control qubit, and $Z^{o_2}$ represents a $\sigma_Z$ gate acting on the $o_2$ qubit. This phase oracle acts as $\hat{O}'_{\vec{x}^{\rm b},\vec{y}^{\rm b}}|i\rangle_n|00\rangle_o= (-1)^{x^{\rm b}_i\oplus y^{\rm b}_i}|i\rangle_n|00\rangle_o$, and the quantum counting scheme counts the number of indexes $i$'s such that $x^{\rm b}_i\oplus y^{\rm b}_i=1$, returning $\frac{d}{N}$ with the same communication complexity as for estimating the correlation.

This result provides a quantum solution to the widely studied gap-Hamming problem in theoretical computer science~\cite{indyk2003tight,chakrabarti2012optimal}. Multiple proofs conclude that it is impossible for a classical protocol to output the Hamming distance $d$ within $\sqrt{N}$ using less than $O(N)$ bits of communication~\cite{hadar2019communication,sherstov2012communication,chakrabarti2012optimal}. By setting $\epsilon = \frac{1}{\sqrt{N}}$, our quantum scheme performs the estimation using $O(\sqrt{N}\log_2(N))$ qubits of communication, exhibiting a square-root speedup over  classical algorithms. As  estimating the Hamming distance under communication constraints has applications in database searching~\cite{indyk2003tight}, networking~\cite{akella2003detecting}, and streaming algorithms~\cite{chakrabarti2010near}, the quantum algorithm can be embedded into various practical classical protocols. 

\paragraph*{Least-square fitting.}
When machine learning models are used to predict the central value of Gaussian distributed continuous variables, the common setting is a linear output layer $f(\boldsymbol{x_i},\boldsymbol{\lambda})=\lambda_0+\vec{\lambda}\cdot\vec{x}=\boldsymbol{\lambda}^T\boldsymbol{x}$ [where $\boldsymbol{x_i}\equiv (1,x_{i,1},\cdots ,x_{i,M-1})^T$ and $\boldsymbol{\lambda}\equiv(\lambda_0,\lambda_1,\cdots ,\lambda_{M-1})^T$] that performs the least-square fitting. The model fitting is reduced to solving a linear least-square problem $\boldsymbol{X}\boldsymbol{\lambda}=\boldsymbol{y}$, where $\boldsymbol{X}\equiv ( \boldsymbol{x}_1,\cdots ,\boldsymbol{x}_N)^T$ is an $N\times M$ matrix belonging to Alice and $\boldsymbol{y}$ is Bob's $N\times 1$ column vector, both of which have real-number components. The goal is to estimate $\hat{\boldsymbol{\lambda}}$ with standard error $\epsilon$ using minimal communications. Here we assume $M\ll N$, as the number of model parameters/attributes is usually much smaller than the number of data points to avoid over-fitting.

The least-square solution of the equation is $\boldsymbol{\lambda}=(\boldsymbol{X}^T\boldsymbol{X})^{-1}\boldsymbol{X}^T\boldsymbol{y}=\frac{1}{N}(N\boldsymbol{X}^{\dagger})\boldsymbol{y}$, where $\boldsymbol{X}^{\dagger}$ is the Moore-Penrose pseudoinverse of $\boldsymbol{X}$; and $N\boldsymbol{X}^{\dagger}$ does not scale with $N$.  As $N\boldsymbol{X}^{\dagger}$ can be computed by Alice locally, only the calculation of $\frac{1}{N}(N\boldsymbol{X}^{\dagger})\boldsymbol{y}$ involves communication. 
The  $j$th component of $\lambda$ can be represented by correlations (inner product) $\lambda_j=\frac1N\sum_i (NX^{\dagger}_{ji}) y_i,    j =0, \cdots, M-1$, which can be calculated by expanding the real numbers as binary floating point numbers. For example, following the IEEE 754 standard~\cite{8766229}, 
each  $N{X}^{\dagger}_{ji}$ and $y_i$ can be written as binary floating point numbers: $NX_{ji}^\dagger\equiv\sum_{k=0}^{\infty}  2^{u-k} x_{ji}^{{\rm b}k}$, $y_{i}\equiv\sum_{k=0}^{\infty}  2^{v-k} y_{i}^{{\rm b}k}$ , where $u$ and $v$ are the highest digit of the elements of $NX_{ji}^\dagger$ and $y_i$, and $x_{ji}^{{\rm b}k}$ and $y_{i}^{{\rm b}k}$ are the $k$th digit, respectively. Then $\lambda_j$ can be written as:
\begin{equation}
\begin{aligned}
\lambda_j &= \frac{1}{N}\sum_{r =0}^{\infty}2^{u+v-r}\sum_{k=0}^r\sum_{i=1}^{N}x_{ji}^{{\rm b}k}y_{i}^{{\rm b}(r-k)}\\
&=2^{u+v}\sum_{r =0}^{\infty}2^{-r}(r+1)f_{jr}.
\end{aligned}
\label{eq:lambda}
\end{equation}
As $x_{ji}^{{\rm b}k}$ and $y_{i}^{{\rm b}k}$ are binary quantity, the inner product $f_{jr}=\frac{1}{N(r+1)}\sum_{k=0}^r\sum_{i=1}^{N}x_{ji}^{{\rm b}k}y_{i}^{{\rm b}(r-k)}$ can be directly estimated by our quantum scheme. The overall communication complexity is $\mathcal C=\sum_{j=1}^{M}\sum_{r=0}^\infty2\frac{\log_2(N)}{\epsilon_{jr}}$,
where $\epsilon_{jr}$ is the standard deviation error of $f_{jr}$. The infinite series in $r$ is cut off according to the target accuracy $\epsilon$ of each component $\lambda_j$, setting $\epsilon_{jr}$ to   $\epsilon_{jr}=\epsilon\frac{0.449}{2^{u+v}(r+1)^{2/3}}2^{\frac{2}{3}r}$. 
If $r$ is large enough so that $\epsilon_{jr}>1$, the quantum algorithm is no longer pertinent, as the number $t$ of ancilla qubits in the quantum phase estimation algorithm drops to less than one, since $\epsilon_{jr} = 2^{-t+1}$.
In that case, $f_{jr}$  can be simply dropped because these $f_{jr}$ terms are multiplied by $2^{-r}$  in Eq.~\eqref{eq:lambda}, they do not contribute substantially to the total error of $\lambda_j$. 
Rewriting $\mathcal C$ in terms of the condition number $\kappa=\parallel{\bf A}^{-1}\parallel_\infty \parallel{\bf A}\parallel_\infty$ of the matrix ${\bf A} = \frac{1}{N}\boldsymbol{X}^T\boldsymbol{X}$ gives 
\begin{equation}
    \mathcal C = 11.026\times2^{v+1}2^uM\frac{\log_2(N)}{\epsilon} = O\left(\frac{M\kappa \log_2(N)}{\epsilon}\right),
    \label{def:C_LSF}
\end{equation} where the absolute magnitude of $2^{v+u}$ in $\mathcal C$ is on the same order of $\frac{\kappa |y|_\infty}{\parallel X\parallel_\infty}$ (see SM section~\uppercase\expandafter{\romannumeral3} for details). The total number of oracle queries is $\mathcal C_{\rm comp}=\frac{M\kappa}{\epsilon}$.

An HHL-based quantum algorithm has been previously developed for data fitting without the communication bottleneck~\cite{wiebe2012quantum}. The algorithm  produces a quantum state $|\boldsymbol{\lambda}\rangle \equiv \sum_j \lambda_j |j\rangle$ with $O(\frac{s^3\kappa^6}{\epsilon}\log_2(N))$ computational complexity, where $0\le s\le 1$ is the sparseness of the matrix $A$. As explained above, this method is, however, inefficient in extracting classical data from the quantum states. 
In the communication-restricted scenario, the HHL-based algorithm requires sharing $O(\frac{\kappa^5M^2}{\epsilon^2}\log_2(N))$ qubits.
For a target statistical precision $\epsilon =1/{\sqrt{N}}$, the quantum counting scheme again obtains a square-root speedup from $O(N)$ to $O(\sqrt{N}\log_2(N))$ compared to the classical theoretical limit. 

\begin{figure}[t]
\centering
\includegraphics[width=\linewidth]{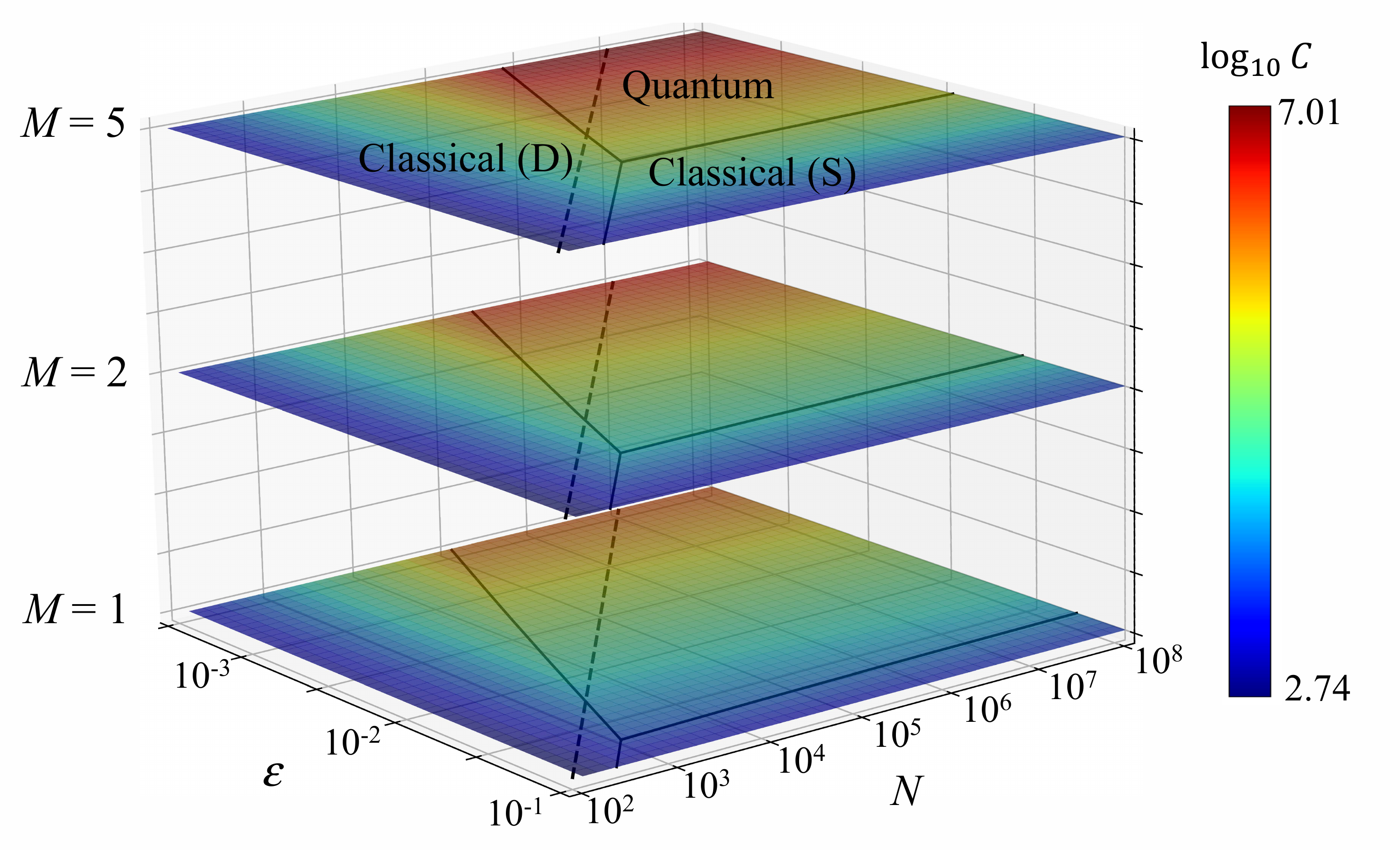}
\caption{Communication complexity phase diagram of the quantum counting algorithm, deterministic, and stochastic classical algorithms in parameter space of $N$, $\epsilon$, and $M$. Without loss of generality, we assume that both $\vec{x}$ and $y$ are normalized, and different components of $\vec{x}$ are \textit{iid}. The color map represents the minimal communication complexity of the three algorithms in the logarithmic scale. Black lines divide the space into three regions denoted as Classical (D), Classical (S), and Quantum, representing the region where deterministic classical, stochastic classical, and quantum counting algorithm has the smallest communication complexity. The black dashed line in each layer indicates the statistical variance $\epsilon=1/\sqrt{N}$.}
\label{fig:phase}
\end{figure}

After demonstrating that the quantum counting algorithm can reduce the communication complexity to $N$, we numerically assess the practical conditions when the quantum algorithm shows an advantage compared to classical algorithms (Fig.~\ref{fig:phase}). In general, the quantum algorithm starts showing an advantage when $N\ge 10^3 \sim 10^4$, which is a reasonable range in fitting problems. The quantum advantage requires $\epsilon$ to be in an intermediate level: too-small or too-large $\epsilon$ make deterministic or stochastic classical algorithm to have a lower communication complexity. 

The quality of a fitted model can be characterized by the mean square error $E \equiv\frac{1}{N} (\boldsymbol{y}-\boldsymbol{X}\hat{\boldsymbol{\lambda}})^2=\frac{1}{N}(\boldsymbol{y}^2+\hat{\boldsymbol{y}}^2-2\boldsymbol{y}^T\hat{\boldsymbol{y}})$. Only the calculation of $\frac{1}{N}\boldsymbol{y}^T\hat{\boldsymbol{y}}$ involves communication, which can again be realized through the correlation estimation scheme, requiring $O(\frac{\log_2(N)}{\epsilon})$-qubit communication. A summary of the communication complexity of different schemes is  presented in Table~\ref{table:complexity}. We also show the computational complexity measured by the number of oracle calls, but leave its implications  to future work.

The applications of the quantum-counting based algorithm are not restricted to fitting linear functions, as a general function of $\vec{x}$ can be expanded as a linear combination of a series of basis functions $y=\sum_j\lambda_jf_j(\vec{x})$. The matrix $F_{ij} \equiv f_j(\vec{x}_i)$ can be computed locally, and the problem is then reduced to the linear fitting problem $\boldsymbol{F}\boldsymbol{\lambda}=\boldsymbol{y}$.  Furthermore, the scheme can be used as the common linear output layer of neural network in high-expressivity machine learning models~\cite{lecun2015deep}.

\paragraph*{Softmax classifier.}  
Besides fitting continuous data, the quantum counting scheme can also be used for fitting discrete labels (classification). A common output layer of classification models is the softmax classifier. The basic scenario is that the data of Bob $y_i$ has discrete possible values in a set of classes $Y=\{c_1,c_2,\cdots , c_q\}$. 
The model outputs the probabilities for a given data point $\vec{x}$ to be in each class $P(y=c_j|\boldsymbol{x},\boldsymbol{\Lambda})$ with ansatz $P(y=c_j|\boldsymbol{x},\boldsymbol{\Lambda}) = \frac{e^{\boldsymbol{\lambda_j}^T\boldsymbol{x}}}{\sum_l e^{\boldsymbol{\lambda_l}^T\boldsymbol{x}}}$, where the coefficient matrix is $\boldsymbol{\Lambda} \equiv (\boldsymbol{\lambda}_0, \cdots , \boldsymbol{\lambda}_q) $. 
We define the cross-entropy loss function $L(\boldsymbol{\Lambda}) \equiv -\sum_{ij}1_{y_i=c_j}\log_2{P(y_i=c_j|\boldsymbol{x}_i,\boldsymbol{\Lambda})}$ to be minimized, where $1_{y=c_j}$ is a 1 when $y=c_j$ and 0 otherwise.
$\boldsymbol{\hat{\lambda}}$ can be obtained from a set of equations:
\begin{equation}
\sum_{i=1}^{N}\frac{\boldsymbol{x}_ie^{\boldsymbol{\hat{\lambda}_j^T\boldsymbol{x}_i}}}{\sum_{k=1}^{q}e^{\boldsymbol{\hat{\lambda}}_k^T\boldsymbol{x}_i}} = 
\sum_{i=1}^N 1_{y_i=c_j} \boldsymbol{x}_i,  \qquad   j = 1, 2, \cdots , q.
\end{equation}
The equation right-hand-sides  can be estimated as inner product between  $1_{y=c_j}$ and the vector $\boldsymbol{x}$ following our previous scheme, with communication complexity $\mathcal C = O(\frac{qM\log_2(N)}{\epsilon})$ (see SM section~\uppercase\expandafter{\romannumeral4} for details). As the left-hand-side of the equations does not involve  $y$, the equations can  be solved without any further communication. We note that logistic regression for the 2-class classification problems can be derived as a special case of the softmax regression scheme with $q=2$.

We can further quantify the communication complexity of evaluating the quality of a fitted classifier. The quality can be determined 
by comparing the model outputs $\hat{y}_i= {\rm argmax}_{c_j} P(y_i=c_j|x_i,\boldsymbol{\Lambda})$ and labels $y_i$ on the training or testing dataset. Alice and Bob encode $\hat{y}_i$ and $y_i$ into $Nq$-bit strings $\hat{b}_{ij}\equiv 1_{\hat{y}_i=c_j}$ and $b_{ij}\equiv 1_{y_i=c_j}$, respectively. Then the correctness of the model can be determined by  estimating the Hamming distance $d$ between $\hat{b}$ and $b$ as $1-\frac{d}{2N}$ (as each error in classification contributes 2-bit difference). 
The communication complexity is $\mathcal C=O(\frac{\log_2{(Nq)}}{\epsilon})$, showing no dependence on dimension $M$ and insensitive dependence on the number of classes $q$.

\paragraph*{Conclusion and Outlook} -
In this work, we developed a quantum counting-based scheme that performs distributed least-square fitting and softmax regression  with a communication complexity  $O(\frac{\log_2(N)}{\epsilon})$, a square-root improvement over classical algorithms. The quantum advantage comes from reduced communication requirements in estimating the correlation and Hamming distance of distributed data, which is achieved by encoding them in the phases of a superposition state, a unique attribute of quantum systems. The quantum phase estimation algorithm then extracts the phase in $O(\frac{1}{\epsilon})$ iterations compared to $O(\frac{1}{\epsilon^2})$  samplings in classical random algorithms. Some previous quantum schemes~\cite{wiebe2012quantum,wang2017quantum,fanizza2020beyond} encode the information in the weight of superposition. As extracting the superposition weight by state tomography also requires $O(\frac{1}{\epsilon^2})$ repetitions of state preparation and measurements, these methods do not show significant advantage in deriving classical fitting parameters compared to classical schemes.

We expect the advantage of our method to benefit  several typical  scenarios of distributed data fitting, demonstrating the benefits of our results, for example: 1. different attributes of the training dataset are collected by different machines with limited communication bandwidth, such as weather data from distant positions or various instruments~\cite{haupt2015big}.
2. The training data is distributed into different workstations for data processing to utilize more CPUs and memory for large-scale machine learning~\cite{gheisari2017survey}. 3. The training process requires privacy preserving and data masking~\cite{al2019privacy}. In our quantum scheme, neither Alice nor Bob can determine the other party's attributes of a specific data point, as only the statistical average is encoded in the phase during communication.  

\ $Acknowledgements -$
We thank Prof. Isaac Chuang for insightful comments. This work was supported by HRI-US, NSF DMR-1923976, NSF DMR-1923929 and NSF CMMI-1922206. The calculations in this work were performed in part on the Texas Advanced Computing Center (TACC) and MIT engaging cluster.

\bibliography{bibliography}

\end{document}